\documentclass[twocolumn]{aastex631}

\usepackage{graphicx}
\usepackage{gensymb}

\newcommand{\htwo}{H$_2$}
\newcommand{\loghtwo}{log $N$(H$_2$)}
\newcommand{\loghone}{log $N_\mathrm{HI}$}
\newcommand{\vlsr}{$v_\mathrm{LSR}$}
\newcommand{\kms}{km s$^{-1}$}

\newcommand{\cm}{cm$^{-2}$}

\newcommand{\fmol}{$f_\mathrm{H_2}$}

\defcitealias{savage2017}{S17}

\received{November 10, 2021}
\revised{November 21, 2021}
\accepted{November 22, 2021}

\submitjournal{ApJL}

\shorttitle{Molecular Gas within the Milky Way's Nuclear Wind}
\shortauthors{Cashman et al. 2021}
\graphicspath{{./}{}}
\begin{document}

\title{Molecular Gas within the Milky Way's Nuclear Wind}

%TC:ignore
\correspondingauthor{Frances Cashman}
\email{frcashman@stsci.edu, afox@stsci.edu}

\author[0000-0003-4237-3553]{Frances H. Cashman}
\affiliation{Space Telescope Science Institute, 3700 San Martin Drive, Baltimore, MD 21218, USA}

\author[0000-0003-0724-4115]{Andrew J. Fox}
\affiliation{AURA for ESA, Space Telescope Science Institute, 3700 San Martin Drive, Baltimore, MD 21218, USA}

\author[0000-0001-8016-6980]{Blair D. Savage}
\affiliation{Department of Astronomy, University of Wisconsin-Madison, 475 North Charter Street, Madison, WI 53706, USA}

\author[0000-0002-0507-7096]{Bart P. Wakker}
\affiliation{Department of Astronomy, University of Wisconsin-Madison, 475 North Charter Street, Madison, WI 53706, USA}

\author[0000-0002-7955-7359]{Dhanesh Krishnarao}
\affiliation{NSF Astronomy \& Astrophysics Postdoctoral Fellow, Johns Hopkins University, 3400 N. Charles Street, Baltimore, MD 21218, USA}
\affiliation{Department of Physics, Colorado College, 14 East Cache La Poudre Street, Colorado Springs, CO 80903, USA}
\affiliation{Space Telescope Science Institute, 3700 San Martin Drive, Baltimore, MD 21218, USA}

\author[0000-0002-8109-2642]{Robert A. Benjamin}
\affiliation{Department of Physics, University of Wisconsin-Whitewater, 800 West Main Street, Whitewater, WI 53190, USA}

\author[0000-0002-1188-1435]{Philipp Richter}
\affiliation{Institut f\"{u}r Physik und Astronomie, Universit\"{a}t Potsdam, Haus 28, Karl-Liebknecht-Str. 24/25, D-14476, Potsdam, Germany}

\author[0000-0002-6541-869X]{Trisha Ashley}
\affiliation{Space Telescope Science Institute, 3700 San Martin Drive, Baltimore, MD 21218, USA}

\author[0000-0003-1892-4423]{Edward B. Jenkins}
\affiliation{Department of Astrophysical Sciences, Princeton University, Princeton, NJ 08544-1001, USA}

\author[0000-0002-6050-2008]{Felix J. Lockman}
\affiliation{Green Bank Observatory, P.O. Box 2, Rt. 28/92, Green Bank, WV 24944, USA}

\author[0000-0002-3120-7173]{Rongmon Bordoloi}
\affiliation{Department of Physics, North Carolina State University, 421 Riddick Hall, Raleigh, NC 27695-8202, USA}

\author{Tae-Sun Kim}
\affiliation{Department of Astronomy, University of Wisconsin-Madison, 475 North Charter Street, Madison, WI 53706, USA}
%TC:endignore

%TC:ignore
\begin{abstract}
We report the first direct detection of molecular hydrogen associated with the Galactic nuclear wind. The Far-Ultraviolet Spectroscopic Explorer spectrum of LS 4825, a B1 Ib--II star at $l,b$ = 1.67\degree,$-$6.63\degree\ lying $d$ = 9.9$^{+1.4}_{-0.8}$ kpc from the Sun, $\sim$1 kpc below the Galactic plane near the Galactic Center, shows two high-velocity \htwo\ components at \vlsr\ = $-79$ and $-108$ \kms.  
In contrast, the FUSE spectrum of the nearby ($\sim$0.6\degree\ away) foreground star HD 167402 at $d$=4.9$^{+0.8}_{-0.7}$ kpc reveals no \htwo\ absorption at these velocities. 
Over 60 lines of \htwo\ from rotational levels $J$ = 0 to 5 are identified in the high-velocity clouds. For the \vlsr\ = $-79$ \kms\ cloud we measure total \loghtwo\ $\geq$ 16.75 \cm, molecular fraction \fmol\ $\geq$ 0.8\%, and $T_{01}$ $\geq$ 97 and $T_{25}$ $\leq$ 439 K for the ground- and excited-state rotational excitation temperatures. At \vlsr\ = $-108$ \kms, we measure \loghtwo\ = 16.13 $\pm$ 0.10 \cm, \fmol\ $\geq$ 0.5\%, and $T_{01}$ = 77$^{+34}_{-18}$ and $T_{25}$ = 1092$^{+149}_{-117}$ K, for which the excited-state ortho- to para-\htwo\ is 1.0$^{+0.3}_{-0.1}$, much less than the equilibrium value of 3 expected for gas at this temperature. This non-equilibrium ratio suggests that the $-108$ \kms\ cloud has been recently excited and has not yet had time to equilibrate. As the LS 4825 sight line passes close by a tilted section of the Galactic disk, we propose that we are probing a boundary region where the nuclear wind is removing gas from the disk.
\end{abstract}
%TC:endignore

%TC:ignore
%TC:endignore

\section{Introduction} \label{sec:intro}

\begin{figure*}[ht!]
    \centering
    \includegraphics[scale=0.85]{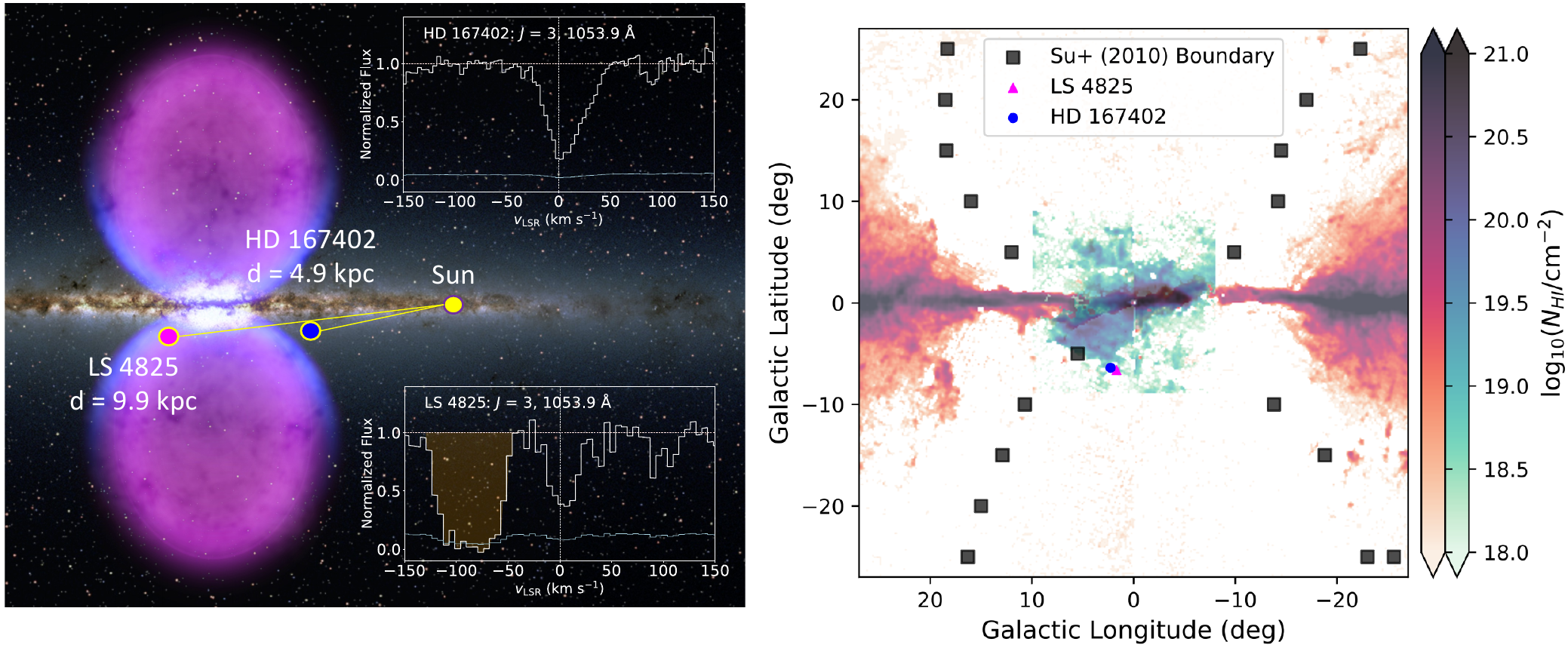}
    \caption{(Left) Edge-on depiction of the Galaxy, with the positions of LS 4825 and HD 167402 marked at their Gaia EDR3 distances of 9.9 and 4.9 kpc, respectively, from the Sun. The purple symmetric lobes show the Fermi Bubbles (image adapted from NASA's Goddard Space Flight Center). The blue edges of the lobes show X-ray emission from ROSAT \citep{bland2003}. The inset panels show the absorption profiles of \htwo\ $J$ = 3 1053.9 {\AA}, one of over 60 lines of \htwo\ detected in the FUSE spectra. The vertical line at 0 \kms\ marks absorption associated with the MW. The strong absorption shaded in orange near $-100$ \kms\ in the background spectrum of LS 4825 is not present in the foreground spectrum of HD 167402, thus bracketing the absorbing region to $4.9 < d < 9.9$ kpc.  (Right) 21 cm \ion{H}{1} emission map from the GASS survey showing the \ion{H}{1} column density at the Galactic tangent points (red scale) from \citet{lockman2016} and at velocities anomalous to circular rotation which track the tilt of the disk near the Galactic Center (green scale: left half from $-110$ to $-70$ \kms, right half from $+70$ to $+110$ \kms). The black squares denote the gamma-ray boundaries of the Fermi bubbles from \citet{su2010}. The foreground FUSE sight line toward HD 167402 at $d$ = 4.9 kpc is marked with a blue circle and the background sight line toward LS 4825 at $d$ = 9.9 kpc is marked with a magenta triangle.}
    \label{fig:fb}
\end{figure*}

The Galactic Center (GC) is host to the Milky Way's (MW) nuclear wind, powered by 
the supermassive black hole Sagittarius A$^{*}$ and surrounding regions of intense star formation. Evidence for the nuclear activity comes from multiple sources, the most notable being the Fermi \citep{su2010, ackermann2014} and eROSITA Bubbles \citep{bland2003, predehl2020}, giant gamma- and X-ray lobes extending $\sim$10 kpc above and below the Galactic plane (see Figure \ref{fig:fb}), which outline the present boundaries of the wind \citep[see also][]{sofue2021}. 
Additional evidence is provided by an \ion{H}{1} outflow, seen in the form of several hundred hydrogen 21 cm clouds \citep{mcclure2013, diteodoro2018, lockman2020} detected at low latitude within the Fermi Bubbles.  
Finally, ultraviolet (UV) absorption-line studies reveal high-velocity absorption in low-ionization (e.g. \ion{C}{1}, \ion{O}{1}, \ion{S}{2}) and high-ionization (e.g. \ion{C}{4}, \ion{Si}{4}) species in sight lines through the Fermi Bubbles. 
These UV absorbers trace outflowing gas that spans an extremely large range in physical conditions \citep{keeney2006, zech2008, fox2015, bordoloi2017, savage2017, karim2018, ashley2020}. Nuclear outflow is also detected in H$\alpha$ emission (see \citealt{krishnarao2020L}). 

Together these observations suggest that a multiphase nuclear outflow exists, with neutral, warm-ionized, and highly ionized components. But until now, very little information on \emph{molecular} gas in the nuclear outflow has existed, with the only information being from two CO emission-line clouds reported by \citet{diteodoro2020}, and no detection in near-IR \htwo\ emission lines \citep{fox2021}. ALMA observations toward J1744-3116 show millimeter-wave molecular absorption (HCO$^+$, HCN, CS) at velocities arising inside the Galactic bulge near the GC \citep{liszt2018}.

High-velocity molecular gas is rarely detected \emph{anywhere} in the Galactic halo, not just the GC. The only UV absorption-line detections of H$_2$ in high-velocity clouds (HVCs) are by \citet{richter1999, richter2001}, \citet{sembach2001}, and \citet{wakker2006}, 
with low molecular fractions $f$(H$_2$)$\sim10^{-6}-10^{-2}$. 

In this paper, we present the first detection of H$_2$ associated with the MW nuclear wind from analysis of archival spectra of two closely spaced low-latitude stars, LS 4825 and HD 167402, observed by the Far-Ultraviolet Spectroscopic Explorer \citep[FUSE;][]{moos2000}. At low latitudes close to the GC, tracing the nuclear outflow is complicated by irregularities in the shape of the Galactic disk \citep{liszt1980,krishnarao2020}, which make the disk/wind separation nontrivial. The sight lines lie in close proximity to a warped portion of the Galactic disk, indicating that we may be probing an interface between the disk and the nuclear wind. 

\section{\htwo\ Data and Measurements} \label{sec:meas}
LS 4825 is a B1 Ib--II blue supergiant with extinction $E(B-V)$ = 0.24 (\citealt{savage2017}, hereafter \citetalias{savage2017}) located at $l,b$ = 1.67\degree,$-$6.63\degree. HD 167402, a B0 Ib supergiant with $E(B-V)$ = 0.23 \citep{shull2019}, lies $\sim$0.6\degree\ away at a maximum projected distance of $\sim$50 pc, at $l,b$ = 2.26\degree,$-$6.39\degree. \citet{ryans1997} measured a spectroscopic distance of 21$\pm$5 kpc for LS 4825, though recent Gaia EDR3 astrometry measurements \citep{bailer2021} place LS 4825 at a much closer distance of 9.9$^{+1.4}_{-0.8}$ kpc. HD 167402 has multiple estimates of its spectrophotometric distance, with \citetalias{savage2017} reporting 7.0$\pm$1.7 kpc and \citet{shull2019} finding 7.6 kpc
(see also \citealt{shull2021}). Gaia EDR3 reports $d$ = 4.9$^{+0.8}_{-0.7}$ kpc for HD 167402. 
We adopt the Gaia distances in our analysis. Through comparative analysis of the foreground HD 167402 and background LS 4825 sight lines, we isolate absorption
from the interval $5 \lesssim d \lesssim 10$ kpc assuming that the ISM coherence length is larger than the projected separation.

FUSE observations of LS 4825 and HD 167402 were performed on 2000 August 29 under program ID P101 (PI: K. Sembach). The raw spectra were obtained from the FUSE archive, and the \texttt{CalFUSE} pipeline (v3.2.1, \citealt{dixon2009}) was used to extract the spectra. The SiC channels ($\lambda < 1000$\,\AA) show complex, overlapping absorption and low signal-to-noise (S/N) ratios and were not used for the analysis. Instead we focus on the spectra from the LiF1 and LiF2 channels, which have S/N$\sim$9$-$13 per resolution element and a velocity resolution of 20\,\kms\ (FWHM). The data were binned by three pixels 
for the fitting analysis. A detailed explanation of the refinements to the \texttt{CalFUSE} data reduction procedures can be found in \citet{wakker2003} and \citet{wakker2006}.

\begin{figure*}[ht!]
    \centering
    \includegraphics[scale=1.4]{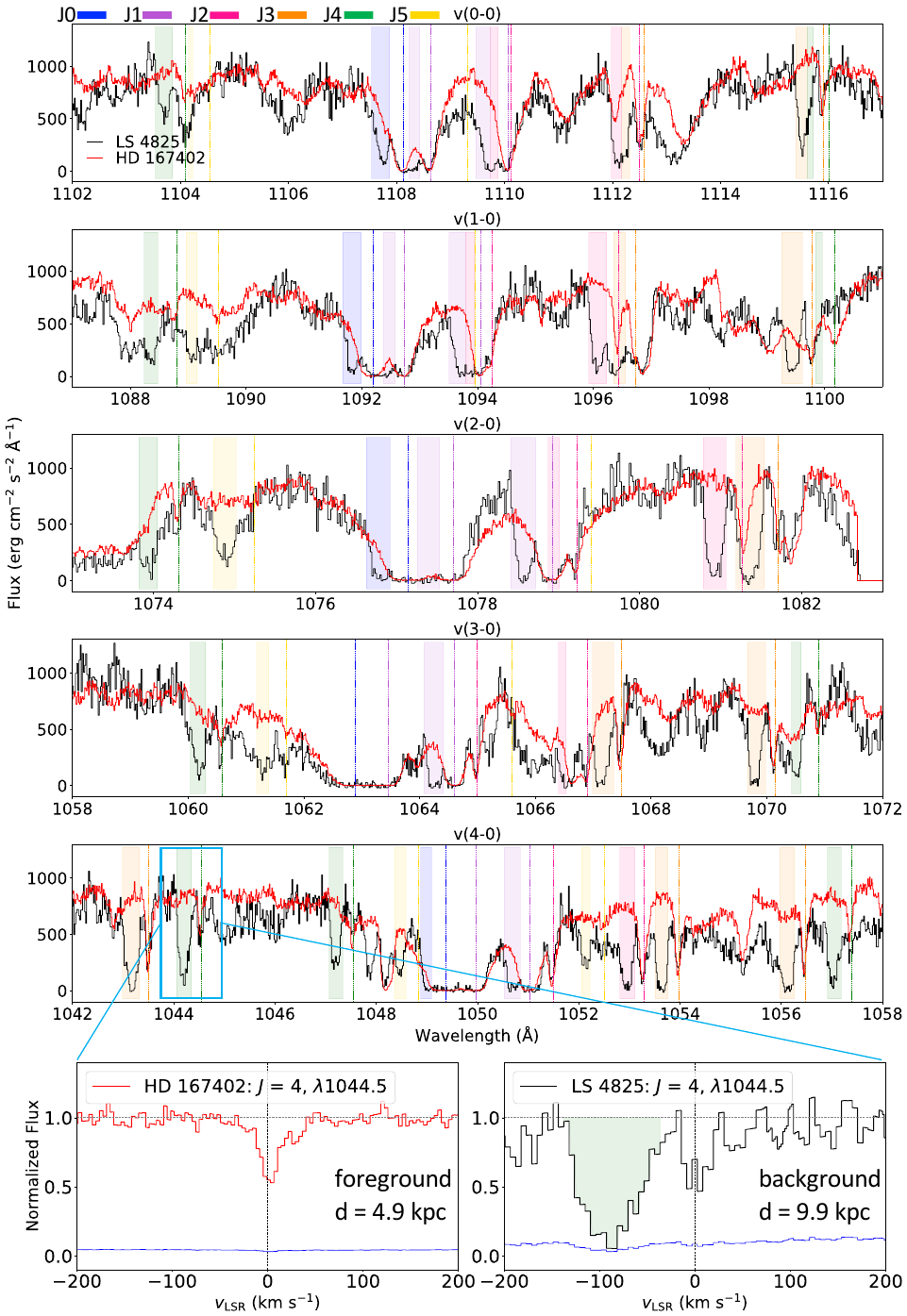}
    \caption{
    FUSE spectra of LS 4825 (background star; black) and HD 167402 (foreground star; red) showing five vibrational \htwo\ bands, each of which has rotational substructure. 
    The LS 4825 spectrum has been multiplied by the average flux ratio of the two stars in the corresponding region to facilitate the comparison. 
    Absorption associated with MW for $J$-levels 0, 1, 2, 3, 4, and 5 are marked in the panels with blue, purple, pink, orange, green, and yellow vertical lines, respectively. The shaded bands mark the region of the HVCs in the background spectrum. The bottom two side-by-side panels show separate radial velocity plots of the normalized data for the $J$ = 4 1044 {\AA} transition to illustrate the difference between the background and foreground spectra, where the vertical line at 0 \kms\ marks absorption associated with the MW, and the 1$\sigma$ error in the normalized flux is shown in blue. The HD 167402 (foreground) spectrum only shows \htwo\ absorption associated with the MW near 0 \kms. The LS 4825 (background) spectrum shows additional strong multi-component \htwo\ absorption centered near $-100$ \kms.}
    \label{fig:comp}
\end{figure*}

The FUSE spectrum of LS 4825 shows high-velocity \htwo\ absorption in over 60 distinct lines from the rotational levels $J$ = 0, 1, 2, 3, 4, and 5 (see Figure \ref{fig:comp}). Two components are seen at \vlsr\ = $-79.4\pm1.5$ and $-107.6\pm1.3$ \kms. Both components have high deviation velocities \citep{wakker1991}, thus we label both as HVCs even though the $-79$ \kms\ component is below the commonly used HVC threshold of $|v|=90$ \kms.
The spectrum of LS 4825 is complex due to its spectral type (B1 Ib$-$II), and the stellar continuum placement was guided by reference to the behavior of the FUSE spectrum of the comparison star HD 58510 (FUSE program ID P102; PI: K. Sembach). HD 58510 has the identical spectral type \citepalias{savage2017} and tracks the continuum of LS 4825 closely, with similar zero-velocity \htwo\ absorption but without the high-velocity \htwo\ absorption. The continua were normalized in local regions of interest using \texttt{linetools} \citep{prochaska2017}.

We used the \texttt{VPFIT} (v12.2, \citealt{carswell2014}) Voigt-profile fitting software to simultaneously fit the individual HVC components for the $J$ = 4 and 5 transitions, as these transitions are unsaturated.  
Since the two HVC components in the lower-$J$ levels ($J < 4$) show significant blending, we assume the velocity centroid and $b$-value derived from $J$=4 and 5 (see Table~\ref{tab:colden}) apply to the $J<4$ levels.  
There is some evidence that $b$-values of interstellar \htwo\ may vary as a function of $J$ \citep{spitzer1976, jenkins1997},
but because of saturation we are unable to determine whether this is the case in our dataset. \citet{jensen2010} performed a study on a sample of 22 Galactic sight lines to determine the effects of assuming a uniform $b$-value for $J$ $\geq$ 2 versus independent measurements and overall found \loghtwo\ differences $<$ 0.07 dex for the excited states.
We quantify the effect this may have on our $J < 4$ column densities (including $J$ = 0 and 1) by lowering $b$ by 3 \kms, as any larger reduction in $b$ results in fits that overfit the profile. For weak unsaturated transitions we find that \loghtwo\ at $-79$ and $-108$ \kms\ increases by at most 0.04 dex. 

We denote the resulting total column density for a component of a given $J$-level as a lower limit if the profile of the \emph{weakest} (lowest $f$-value) transition in the set of simultaneously fit lines reaches zero flux.  
\htwo\ lines blended with ISM metal lines were excluded from fitting, as were lines blended with other \htwo\ lines and those in regions of geocoronal emission. All lines of a single $J$-level were visually inspected in velocity space to verify that when progressing from weakest to strongest lines the absorption behaved as expected, and to check for the presence of damping wings on the HVC \htwo\ components. Following continuum reconstruction using the reference star and the adoption of a two-component fit, we find no evidence for HVC damping wings. This is also consistent with the very small differential extinction between the foreground and background sight lines using  $E(B-V)$ values reported in \citetalias{savage2017} and \citet{shull2019}, $\Delta E(B-V)$ = 0.01, because this indicates a similar total dust column and therefore similar total hydrogen column in the two directions.  
 
\begin{figure}[ht!]
    \centering
    \includegraphics[scale=1.11]{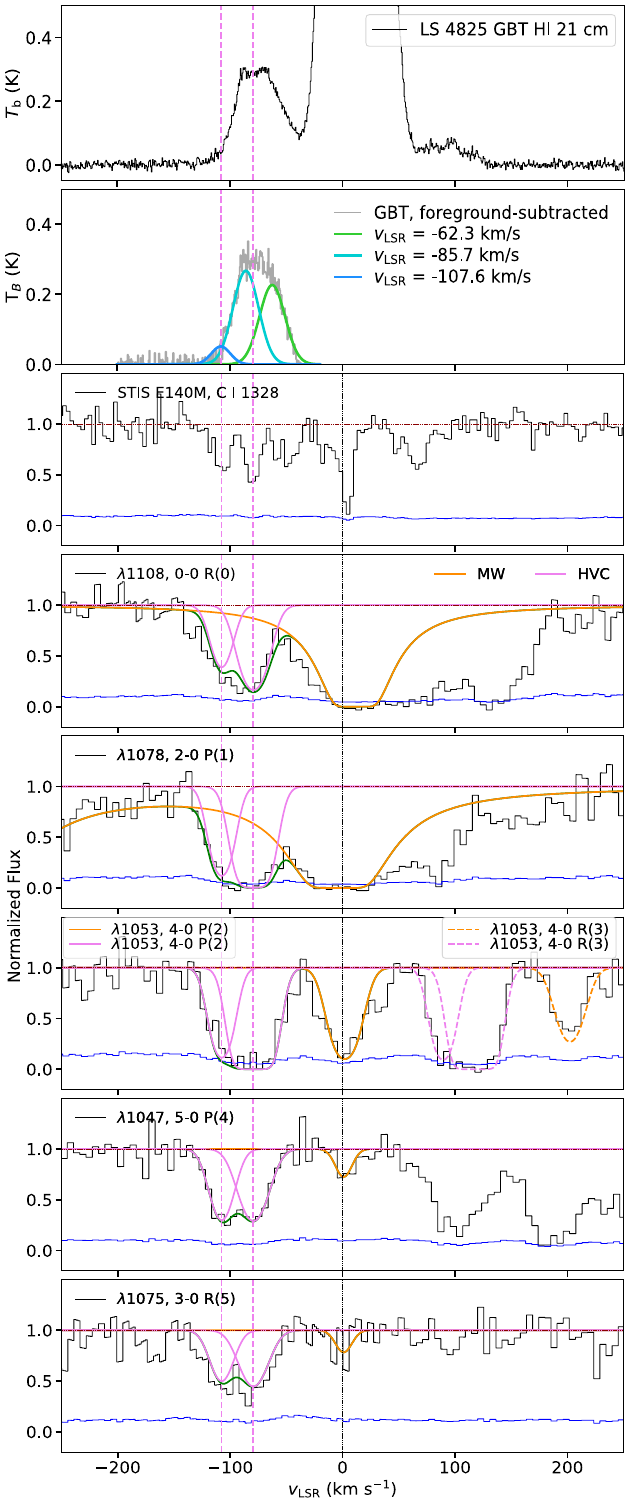}
    \caption{Velocity profiles of \htwo\ absorption lines and \ion{H}{1} emission toward LS 4825. Panel 1: LS 4825 21 cm GBT spectrum from \citetalias{savage2017}. Panel 2: refit to the MW foreground-subtracted spectrum with \loghone\ = 19.14$\pm$0.16 at $-86$ \kms\ and $\leq$ 18.77 at $-108$ \kms. Panel 3: closely aligned \htwo\ from this work with \ion{C}{1} from \citetalias{savage2017}. Panels 4--8: a sample of \htwo\ absorption lines for rotational levels $J$ = 0--5. The normalized flux is shown in black, the continuum level is in red, and the 1$\sigma$ error in the normalized flux is in blue. The vertical line at 0 \kms\ marks the region associated with the MW. The solid orange and magenta curves are Voigt profile fits to the \htwo\ absorption features for the MW and the HVCs, respectively. In all panels, the two vertical dashed magenta lines indicate the velocities of the HVC components.}
    \label{fig:vpfit}
\end{figure}

We employed a slightly different procedure to fit the HVC ground state $J$ = 0, 1 components due to overlapping with strong Milky Way (MW) foreground \htwo\ absorption. For these lines we fit a Voigt profile to the MW \htwo\ component with fixed $b$=5.5 \kms\ and adjusted its column density interactively until the profile best fit the damping wings for all selected lines of the given $J$-level. The choice of $b$=5.5\,\kms\ was motivated by published high-resolution (FWHM = 6.6 \kms) STIS observations of \ion{C}{1} toward LS 4825 \citepalias{savage2017}, where $b_\mathrm{CI}$=5.4 \kms\ at $T$ $\sim$100 K indicates that the gas is almost fully turbulent. The velocity, $b$-value, and column density of the MW foreground component were then held fixed, and the column density for the HVC components was then determined as described above for the excited $J$-levels. The resulting Voigt profile fits to the \htwo\ lines are shown in Figure \ref{fig:vpfit} and the resulting velocities, $b$-values, and column densities are shown in Table \ref{tab:colden}.  

\begin{figure}[ht!]
    \centering
    \includegraphics[scale=0.50]{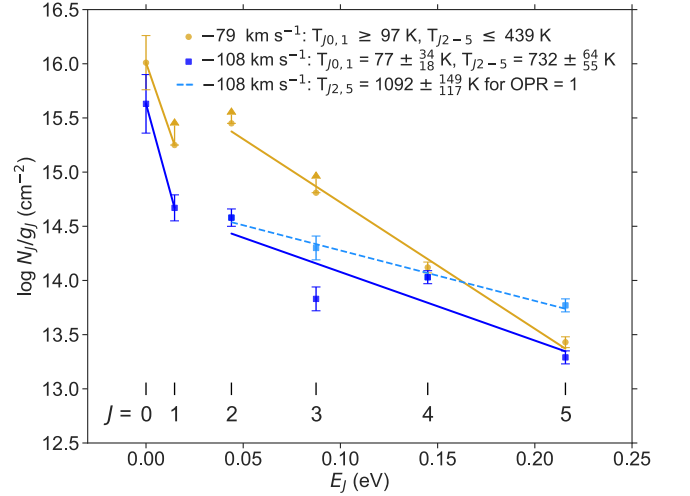}
    \caption{Boltzmann excitation plot for the two high-velocity \htwo\ components toward LS 4825. The \htwo\ column density in each rotation level $N_J$ divided by the level's statistical weight $g_J$ is plotted as a function of excitation energy ($E_J$) for each cloud, for $J$=0--5. The HVCs at $-79$ and $-108$ \kms\ are indicated with gold circles and blue squares, respectively. For the cloud at $-79$ \kms, we adopt a two-slope solution to describe the relative population of the ground-state ($J$=0,1) and excited-state levels ($J$=2--5), as indicated in the legend. For the cloud at $-108$ \kms, we also consider a model in which the ortho-to-para ratio (OPR) of the excited-state levels is 1 instead of 3, which would shift log $N_J$/$g_J$ for $J$=3,5 upward to the positions indicated by the light blue squares and for which the light blue dashed line is a better fit to the distribution of points. 
    }
    \label{fig:boltz}
\end{figure}

\begin{table*}[ht!]
\caption{HVC Molecular Absorption-Line Measurements}
\label{tab:colden}
\begin{center}
\begin{tabular}{lcccc}
\hline\hline
Species       & $\lambda_\mathrm{rest}$ & $f^{a}$                 & log $N$(H$_2$)$_{v=-79}^b$  & log $N$(H$_2$)$_{v=-108}^c$ \\
$\nu_{u}$--$\nu_{l}$, R/P($J_{l}$)  & ({\AA})                 &                         & ($N$ in cm$^{-2}$)      & ($N$ in cm$^{-2}$)      \\
\hline
H$_2$ $J = 0$ &                         &                         & 16.01 $\pm$ 0.25 & 15.63 $\pm$ 0.27 \\
$0-0$, R(0)   & 1108.1273               & 1.66 $\times$ 10$^{-3}$ &                  &                  \\
$1-0$, R(0)   & 1092.1952               & 5.78 $\times$ 10$^{-3}$ &                  &                  \\
$2-0$, R(0)   & 1077.1387               & 1.17 $\times$ 10$^{-2}$ &                  &                  \\
\hline
H$_2$ $J = 1$ &                         &                         & $>$ 16.20         & 15.62 $\pm$ 0.12 \\
$1-0$, P(1)   & 1094.0519               & 1.97 $\times$ 10$^{-3}$ &                  &                  \\
$2-0$, P(1)   & 1078.9255               & 3.92 $\times$ 10$^{-3}$ &                  &                  \\
$4-0$, P(1)   & 1051.0325               & 7.60 $\times$ 10$^{-3}$ &                  &                  \\
\hline
H$_2$ $J = 2$ &                         &                         & $>$ 16.15         & 15.28 $\pm$ 0.08 \\
$2-0$, P(2)   & 1081.2660               & 4.70 $\times$ 10$^{-3}$ &                  &                  \\
$3-0$, P(2)   & 1066.9007               & 7.09 $\times$ 10$^{-3}$ &                  &                  \\
$4-0$, P(2)   & 1053.2843               & 9.02 $\times$ 10$^{-3}$ &                  &                  \\
\hline
H$_2$ $J = 3$ &                         &                         & $>$ 16.13         & 15.15 $\pm$ 0.11 \\
$3-0$, P(3)   & 1070.1409               & 7.54 $\times$ 10$^{-3}$ &                  &                  \\
$3-0$, R(3)   & 1067.4786               & 1.00 $\times$ 10$^{-2}$ &                  &                  \\
$5-0$, P(3)   & 1043.5032               & 1.08 $\times$ 10$^{-2}$ &                  &                  \\
$4-0$, R(3)   & 1053.9761               & 1.34 $\times$ 10$^{-2}$ &                  &                  \\
\hline
H$_2$ $J = 4$ &                         &                         & 15.08 $\pm$ 0.05 & 14.99 $\pm$ 0.06 \\
$3-0$, P(4)   & 1074.3130               & 7.74 $\times$ 10$^{-3}$ &                  &                  \\
$5-0$, P(4)   & 1047.5519               & 1.10 $\times$ 10$^{-2}$ &                  &                  \\
$4-0$, R(4)   & 1057.3807               & 1.29 $\times$ 10$^{-2}$ &                  &                  \\
$5-0$, R(4)   & 1044.5433               & 1.55 $\times$ 10$^{-2}$ &                  &                  \\
\hline
H$_2$ $J = 5$ &                         &                         & 14.95 $\pm$ 0.05 & 14.81 $\pm$ 0.06 \\
$3-0$, R(5)   & 1075.2441               & 9.28 $\times$ 10$^{-3}$ &                  &                  \\
$5-0$, P(5)   & 1052.4970               & 1.11 $\times$ 10$^{-2}$ &                  &                  \\
\hline
Total log $N$(H$_2$)          &              &                         & $>$ 16.75        & 16.13 $\pm$ 0.10 \\
\hline
\ion{H}{1} (\citetalias{savage2017})$^d$ & 21 cm        & ...     & 19.43 $\pm$ 0.15 & 18.28 $\pm$ 0.15\\
\hline
\ion{H}{1} (this work)$^e$ & 21 cm                  & ...                     & 19.14 $\pm$ 0.16 & $<$ 18.77 \\
\hline
$f_\mathrm{H_2}$ (\citetalias{savage2017})$^f$  &                   &                         & $>$ 0.41\%       & 1.40 $\pm$ 0.26\% \\
\hline
$f_\mathrm{H_2}$ (this work)$^g$  &                   &                         & $>$ 0.81\%       & $>$ 0.46\% \\
\hline\hline              
\end{tabular}\\
\end{center}
\footnotesize
$^a$ The wavelengths and oscillator strengths used by \texttt{VPFIT} are calculated from \citet{bailly2010}. \\
$^b$ The $b$-value from \texttt{VPFIT} for this component is 14.9$\pm$1.7\,\kms. \\ 
$^c$ The $b$-value from \texttt{VPFIT} for this component is 11.7$\pm$1.5\,\kms. \\ 
$^d$ \citetalias{savage2017} determine \loghone\ = 19.43$\pm$0.01 and 18.28$\pm$0.02 for the $-79$ and $-108$ \kms\ components in the 21 cm data from the Green Bank Telescope (GBT, program ID: 14B-299). The \loghone\ errors include a beam smearing error of $\pm$0.15 since we are combining the GBT data (9\arcmin.1 beam) with the FUSE UV measurements (infinitesimal beam) to derive \fmol. \\
$^e$ Revised estimate of log $N_\mathrm{HI}$ from a refit to the GBT LS 4825 spectrum in \citetalias{savage2017} where the MW foreground \htwo\ absorption of HD 167402 has been subtracted. The \loghone\ error for the $v$ = $-79$ \kms\ component includes a beam smearing error of $\pm$0.15. \\
$^f$ The molecular fraction, $f_\mathrm{H_2}$~=~2$N$(\htwo)/[$N$(\ion{H}{1}) + 2$N$(\htwo)], based on \loghone\ from \citetalias{savage2017}. \\
$^g$ Estimate of the molecular fractions derived from the revised multi-component refit to the 21 cm \ion{H}{1} GBT spectrum in \citetalias{savage2017}. \\
\end{table*}

\section{Results} \label{sec:results}

Our most significant result is the discovery of the presence of the high-velocity \htwo\ absorption in two components in the spectrum of the background star LS 4825, centered at $-$79 and $-$108\,\kms. This remains true
irrespective of our detailed findings on column densities and rotational excitations discussed below.
The high-velocity \htwo\ components are not seen in the spectrum of the foreground star HD 167402, indicating that the \htwo\ detected at $-79$ and $-108$ \kms\ in the background star LS 4825 can be bracketed to the range $5 \lesssim d \lesssim 10$ kpc, which corresponds to a $z$-distance of $-0.6$ to $-1.2$ kpc below the GC. The \htwo\ is therefore potentially associated with the southern Fermi Bubble, a region exposed to the Galactic nuclear wind. 

The absence of high-velocity absorption toward HD 167402 was also seen in the STIS E140M and E230M UV spectrum by \citetalias{savage2017}, who reported multiphase gas at high velocities only in the LS 4825 spectrum. Low-ion absorption is seen over the range $-290$ to 94 \kms. Of particular significance is \ion{C}{1} absorption detected at \vlsr\ = $-114.2$, $-100.8$, and $-77.7$ \kms\ in the STIS spectrum \citepalias{savage2017} because \ion{C}{1} acts as a tracer for \htwo\ \citep{ge1997,ge2001}. The combination of our results from \htwo\ detected at similar velocities with those from \citetalias{savage2017} allows us to
construct a more complete picture of the composition, properties, and environment of these cold clouds in the dynamic environment near the GC.

\subsection{The \htwo\ HVC at $-$79 km s$^{-1}$} \label{subsec:hvc79}

By summing over the rotational levels from $J$=0 to 5, we derive a total \htwo\ column density in the $-$79\,\kms\ component of \loghtwo\ $>$ 16.75 \cm. This is a lower limit due to saturation in multiple $J$-levels. \citetalias{savage2017} report \ion{H}{1} 21 cm emission at negative velocities of $-105$, $-89$, and $-70$ \kms\ using a Green Bank Telescope (GBT) spectrum and conclude that the $-70$ \kms\ component with \loghone\ = 19.43 $\pm$ 0.01 is most likely associated with the absorption system near $-78$ \kms. Combining this \ion{H}{1} measurement with our measurement of \loghtwo\, we derive a molecular fraction of $f_\mathrm{H_2}$~=~2$N$(\htwo)/[$N$(\ion{H}{1}) + 2$N$(\htwo)] $\geq$ 4.1 $\times$ 10$^{-3}$, or $\geq$0.41\% (see Table \ref{tab:colden}). We conducted a refit to the LS 4825 GBT spectrum, in which we subtracted the zero-velocity component of the foreground HD 167402 spectrum. This results in two components at $-85.8$ and $-62.3$ \kms\ with \loghone\ = 19.14$\pm$0.16 and 19.07$\pm$0.16 \cm, respectively. We associate the \ion{H}{1} emission near $-86$ \kms\ with the \htwo\ absorption at $-79$ \kms\ and find \fmol\  $\geq$ 0.81\%. Either limit is consistent with the upper end of \fmol = $10^{-2}-10^{-6}$ measured in Galactic HVCs from absorption-line studies (\citealt{richter2001, sembach2001, wakker2006}), but is not as high as \fmol\ $\sim$0.3--0.6 from emission-line CO detections seen in \citet{diteodoro2020}.

We determine limits on the excitation temperatures in the high-velocity molecular gas by fitting a theoretical Boltzmann distribution to the observed population of rotational levels, since our measurements of \loghtwo\ for $J$ = 1, 2, 3 are lower limits. For the $-79$ \kms\ component, the rotational ground states $J$ = 0 and 1 are fit by a slope equivalent to a Boltzmann temperature $T_{01}$ $\geq$ 97 K, whereas the excited rotational levels $J$ = 2$-$5 fit on a slope equivalent to $T_{25}$ $\leq$ 439 K (see Figure \ref{fig:boltz}). 
The adopted two-slope solution may suggest a core-envelope structure, with $T_{01}$ reflecting the temperature in the cooler, shielded interior of the cloud and the $T_{25}$ region tracing warmer gas heated by processes such as UV pumping, \htwo\ formation pumping, and shock excitation \citep{richter2001}. However, these processes could occur within a fairly homogeneous region. 

\citetalias{savage2017} measure a near solar metallicity [S/H] = 0.02 $\pm$ 0.16 in the $-$78\,\kms\ HVC ([S/H] = 0.31 $\pm$ 0.17 using the \ion{H}{1} measurement from the foreground-subtracted GBT spectrum), and also report a subsolar Fe/S ratio, finding [\ion{Fe}{2}/\ion{S}{2}] = $-1.05\pm0.05$, which suggests that Fe in the cloud is locked up in dust grains. 

\subsection{The \htwo\ HVC at $-$108 km s$^{-1}$} \label{subsec:hvc108}

We report \loghtwo = 16.13 $\pm$ 0.10 \cm\ for the $-108$ \kms\ component, 0.6 dex lower than the \htwo\ column at $-79$ \kms. From the MW foreground-subtracted GBT spectrum of LS 4825 we determine a 3$\sigma$ upper limit of \loghone\ $\leq$ 18.77 at $-107.6$ \kms, which is $\sim$0.5 dex higher than \citetalias{savage2017}, who measure \loghone\ = 18.28 $\pm$ 0.02 at $-104.9$ \kms.  Combining the \ion{H}{1} and \htwo\ columns results in \fmol\ $\geq$ 0.46\%, which is also consistent with the upper end of \fmol\ for HVCs in absorption-line studies (see section \ref{subsec:hvc79}). 
The metallicity [S/H] $\geq$ 0.87 is supersolar; and \citetalias{savage2017} measure [\ion{Fe}{2}/\ion{S}{2}] = $-0.99\pm0.09$ for this component, concluding that it is as equally dusty as the $-79$ \kms\ component.

A two-slope solution for the Boltzmann distribution fit to the rotational level populations yields $T_{01}$ = 77$^{+34}_{-18}$ K and $T_{25}$ = 732$^{+64}_{-55}$ K, as determined from the dark blue lines in Figure \ref{fig:boltz}, but is not a good fit for the $J \geq 2$ excited states. However, the up-and-down distribution of the excited states, in which the para states $J$=2,4 lie systematically above the ortho states $J$=3,5 may indicate an ortho-to-para ratio (OPR) that deviates from the canonical (equilibrium) value of 3 for warm, rotationally excited \htwo\ gas (as included in the statistical weight, $g_J$, where $g_J$(para)=2$J$+1 and $g_J$(ortho)=3(2$J$+1)). If we adjust the OPR for $J \geq 2$ to a value of 1 ($^{+0.3}_{-0.1}$) instead of 3, the data points for $J$=3,5 (in light blue) are shifted higher such that all $J \geq 2$ levels fit on a straight line, with a slope that yields $T_{{\rm exc}}$ = 1092$^{+149}_{-117}$ K (see Figure \ref{fig:boltz}). This suggests that the OPR is out of equilibrium and is further discussed in Section \ref{sec:disc}. An OPR=1 provides a good fit to the data for the excited states of the $-108$ \kms\ component, however, we acknowledge that unresolved components with potentially varying $b$-values could impact the measurements of the excited state column densities. These fits represent what is currently capable given the resolution of the data.

\section{Discussion} \label{sec:disc}

The \htwo\ components detected in the FUSE spectrum of LS 4825 ($d$ = 9.9$^{+1.4}_{-0.8}$ kpc) at $-79$ and $-108$ \kms\ are not seen in the spectrum of the foreground star HD 167402 ($d$ = 4.9$^{+0.8}_{-0.7}$ kpc) and therefore trace gas located between the two stars and most likely within the southern Fermi Bubble (see Figure~\ref{fig:fb}).
Our results represent the first detection of high-velocity \htwo\ in the extended Galactic Center environment. We measure total \loghtwo\ $>$ 16.75 and 16.13$\pm$0.10 \cm\ at $-79$ and $-108$ \kms\ respectively, velocities which cannot be explained by circular rotation. We determine respective H$_2$ fractions \fmol\ $\geq$ 0.8\% and \fmol\ $\geq$ 0.5\%.
The low ground-state temperatures we derive for the \htwo\ absorption at $-79$ and $-108$ \kms\ of $T_{0,1}\approx$97 and 77 K confirm that these are indeed cold clouds. The \htwo\ detection augments the metal-line analysis from \citetalias{savage2017}, who reported high-velocity neutral, low-ion, and high-ion absorption toward LS~4825 but not HD~167402.

A natural interpretation of these results, as suggested by \citetalias{savage2017}, is
that the multiphase high-velocity absorbers toward LS 4825 trace a nuclear wind.
This is supported by the very strong high-ion absorption seen along the line of sight \citepalias{savage2017} and the 
abundant multi-wavelength evidence for a wind in this region
\citep{bland2003, su2010, predehl2020, diteodoro2018, lockman2020}. 
In this interpretation, only a small component of the outflow velocity is projected along our line of sight; deprojecting the velocity onto a vertically or radially oriented outflow
implies a high outflow velocity ($\sim$800--1000 \kms), similar to the velocity inferred from other 
UV HVCs in the Fermi Bubbles \citep{fox2015, bordoloi2017}.  

However, multiple structures house molecular and neutral gas in the GC. The Galactic disk near the GC is warped, tilted at $\sim$22\degree\ \citep{liszt1980} with a portion protruding to $b\sim-5$\degree\ between $l$=0--10\degree\ (as shown in Figure~\ref{fig:fb}). 
Recent H$\alpha$ studies at the anomalous velocities that track the tilt of the Galactic disk in this region from \citet{krishnarao2020} predict asymmetric absorption profiles with peak absorption near $-100$ \kms, just as we observe. Even closer to the GC lies the Central Molecular Zone inside $R_G\sim$0.5 kpc. The right panel of Figure \ref{fig:fb} shows that the sight lines lie in close proximity to the edge of the disk and could be probing the boundary where the nuclear wind is passing by the disk since the HVCs and the disk have similar velocities in this inner region.
Irrespective of whether the clouds are entrained in an escaping wind, we see strong evidence of disk-like properties for the molecular components from their high metallicities, including an above-solar [S/H] = 0.31$\pm$0.17 at $-79$ \kms\ and a supersolar [S/H] $\geq$ 0.87 near $-108$ \kms, as well as high dust depletion levels, with [\ion{Fe}{2}/\ion{S}{2}] $\sim-1$ \citepalias{savage2017}.

Both the wind interpretation and the warped-disk interpretation 
are allowed by the data. 
However, the two are not mutually exclusive. We propose a hybrid model in which the high-velocity \htwo\ clouds probe a boundary region where the nuclear wind is passing by the disk and accelerating fragments of gas into the halo. This is supported by the sight line's close proximity to the disk (both spatially and kinematically), as well as unusual OPRs and thermal pressures \citepalias{savage2017}, as we explain below. 

Warm, rotationally excited \htwo\ gas in thermal equilibrium is expected to have an OPR=3, as the excited gas has been warm for long enough to reach equilibrium between the (odd) ortho and (even) para states. As seen in Figure \ref{fig:boltz}, an OPR=3 is a good description of the distribution of excited-state levels for the cloud at $-79$ \kms. However, as 
described in Section \ref{sec:results}, an OPR=3 fails to describe the excited $J$ levels for the $-108$ \kms\ cloud, which instead is much better fit by OPR=1. This deviation from the canonical (equilibrium) value suggests that the observed $J$=2--5 levels at $-108$ \kms\ were only recently pumped into high $J$ states 
(and also to higher $T$) from the ground states ($J$=0 and 1), where OPR$\lesssim$1 is expected \citep{flower1984}. 
The OPR equilibrates very slowly \citep{neufeld1998}, so in this explanation the excited-state gas has not had enough time to adjust to the new environment to reach equilibrium between the ortho and para states, and so preserves a "memory" of the thermal state of gas in a previous epoch. 
We note that an enhanced radiation field within the Fermi Bubbles is \emph{predicted} in the models of \citet{bland2019}, who find an ionizing radiation field of log~$\varphi$ = 6.5 photons \cm\ s$^{-1}$ at $\sim$1 kpc below the GC, and potentially higher in the event of a recent Seyfert flare. The unusual \htwo\ excitation properties we observe are consistent with this. 

In the hybrid scenario, the two \htwo\ clouds $\sim$1 kpc below the GC 
undergo different histories. Both clouds formed much earlier in a quiescent disk environment at $T<100$ K and were swept and/or broken up within the nuclear wind. In this new environment, the outer skin of the clouds was rotationally excited by photons and/or collisions with an ambient medium. Whereas the $-79$ \kms\ cloud has equilibrated to this environment with $T_{\rm exc}\approx$492 K for OPR=3, the $-108$ \kms\ cloud has not, as evidenced by $T_{\rm exc}$=1092 K with an OPR=1 that still reflects the physical conditions of the \htwo\ gas from its prior environment. 

Further support for a wind or hybrid model is provided by the high thermal pressure in the HVCs.
From observations of $N$(\ion{C}{1}), $N$(\ion{C}{1}$^{*}$), and $N$(\ion{C}{1}$^{**}$), \citetalias{savage2017} derive thermal pressures $P/k$ at $-78$, $-101$, and $-114$ \kms\ of $\sim$10$^{4.1}$, $10^{3.8}$, and $10^{5}$\,cm$^{-3}$\,K, respectively. The pressures reported at $-78$ and $-114$ are $\sim$3.2 and 25 times higher than the mean $P/k=10^{3.6\pm0.2}$\,cm$^{-3}$\,K reported in the diffuse ISM in the Galactic disk \citep{jenkins2011}, which suggests that the clouds may have been compressed, perhaps by an outflowing hot wind. The pressure at $-101$ \kms, however, is similar to the mean ISM cloud pressure. Since we measure an \htwo\ velocity centroid in the lower resolution FUSE spectrum at $-108$ \kms, we are unable to make a confirmed association with either of the STIS components at $-101$ and $-114$ \kms, but acknowledge that the range of pressures observed near this velocity are consistent with a hybrid disk-wind environment along this line of sight.

If the clouds are indeed being actively swept out of the disk, perhaps in a biconical outflow from the GC (see \citealt{fox2015, bordoloi2017}), then questions arise on how the cold gas became entrained and will survive 
in this complex and energetic environment.
HVCs are thought to have a finite lifetime against disruptive instabilities as they interact with a surrounding medium \citep{heitsch2009, armillotta2017}.
However, recent theoretical studies have explored the survival of cold gas in hot galactic winds (e.g. \citealt{gronke2020,sparre2020}). They indicate that a cold gas cloud entrains hot gas via cooling-induced pressure gradients, thereby acquiring the mass and momentum of the hot gas. Inclusion of a magnetic field, and in particular, a turbulent magnetic wind, can suppress the cloud-destroying Kelvin-Helmholtz instabilities, allowing the clouds to survive. 
These models show that clouds with $N_\mathrm{H}$ $<$ 10$^{18}$ \cm\ are not predicted to survive due to erosion, whereas those with $N_\mathrm{H}$ $>$ 10$^{18}$ \cm\ are expected to survive and even grow. 

The results of these models offer a plausible explanation for how the low column density \htwo\ components may survive, as comoving within clouds with $N_\mathrm{H}$ $>$ 10$^{18}$ \cm\ could possibly shield the \htwo\ bearing structures, ensuring their survival. If entrained in a wind, the existence of these high-velocity molecular clouds in the dynamic and multiphase GC environment will help to inform continuing research on models of cloud acceleration and survival. However, further detections of molecular gas at more locations within the nuclear wind are needed to test these theories.

\section{Summary} \label{sec:summary}
We have detected two high-velocity molecular hydrogen clouds in the FUSE spectrum of the massive star LS 4825.
The sight line passes $\sim$1 kpc below the GC near the boundary of the Galactic disk, a region where the nuclear wind is thought to blow gas out into the southern Fermi Bubble.
The clouds are not seen in the spectrum of the foreground star HD 167402, lying $\sim$0.6\degree\ away, confirming they are associated with the GC.
We measure total \loghtwo\ $>$ 16.75 and 16.13$\pm$0.10 \cm\ at $-79$ and $-108$ \kms\ respectively, velocities which cannot be explained by circular rotation. We determine respective H$_2$ fractions \fmol\ $\geq$ 0.8\% and \fmol\ $\geq$ 0.5\%. 
For the $-79$ \kms\ cloud we adopt a two-component Boltzmann distribution to explain the rotational level populations, with $T_{01}$ $\geq$ 97 K and $T_{25}$ $\leq$ 439 K. For the $-108$ \kms\ cloud, a two-component Boltzmann distribution yields $T_{01}$ = 77$^{+34}_{-18}$ and $T_{25}$ = 732$^{+64}_{-55}$ K, but is not a good fit for the excited states given the canonical OPR=3. Instead we find that if OPR=1, the excited states lie on a straight line corresponding to $T_{25}$ = 1092$^{+149}_{-117}$ K. 

We considered two possible origins for the high-velocity \htwo\ components: a wind interpretation which traces the projection of an outflow velocity along our line of sight, and a warped-disk interpretation in which the clouds are associated with the tilted portion of the Galactic disk. We 
conclude that the best explanation for the data is
a hybrid model in which the \htwo\ clouds probe a boundary region where the nuclear wind is passing by the disk and accelerating fragments of gas into the halo.
\vspace{5mm}

%TC:ignore
\begin{acknowledgements} 
We thank Enrico Di Teodoro for valuable conversations on the LS 4825 sight line and Stephen McCandliss for helpful conversations about molecular absorption lines. We gratefully acknowledge support from the NASA Astrophysics Data Analysis Program (ADAP) under grant 80NSSC20K0435, {\it 3D Structure of the ISM toward the Galactic Center.} 
The FUSE data were obtained under program P101. FUSE was operated for NASA by the Department of Physics and Astronomy at the Johns Hopkins University. 
The Green Bank Telescope data were obtained under Program GBT14B-299, and the observatory is a facility of the National Science Foundation, operated under a cooperative agreement by Associated Universities, Inc. 
D.K. is supported by an NSF Astronomy and Astrophysics Postdoctoral Fellowship under award AST-2102490.
\end{acknowledgements}
%TC:endignore

Some of the data presented in this paper were obtained from the Mikulski Archive for Space Telescopes (MAST) at the Space Telescope Science Institute. The specific observations analyzed can be accessed via \dataset[10.17909/t9-axmr-t152]{https://doi.org/10.17909/t9-axmr-t152}.

%TC:ignore
\facilities{FUSE, GBT, HST (STIS)}
%TC:endignore

%TC:ignore
\software{%\texttt{astropy} \citep{astropy2018},
          \texttt{linetools} \citep{prochaska2017}},
          \texttt{VPFIT} \citep{carswell2014}
%TC:endignore

%TC:ignore
\bibliography{ref}{}
\bibliographystyle{aasjournal}
%TC:endignore

%% This command is needed to show the entire author+affiliation list when
%% the collaboration and author truncation commands are used.  It has to
%% go at the end of the manuscript.
%\allauthors

%% Include this line if you are using the \added, \replaced, \deleted
%% commands to see a summary list of all changes at the end of the article.
%\listofchanges

\end{document}